\documentclass[aps,prb,reprint,groupedaddress,superscriptaddress,showpacs]{revtex4-1}
\usepackage{graphicx}
\usepackage{bm}        
\usepackage{amssymb}   

\begin{document}


\title{Modulation of a surface plasmon-polariton resonance by sub-terahertz diffracted coherent phonons}


\author{C. Br\"{u}ggemann}
\email[Email: ]{christian.brueggemann@tu-dortmund.de}
\affiliation{Experimentelle Physik 2, Technische Universit\"{a}t Dortmund, D-44227 Dortmund, Germany}

\author{A. V. Akimov}
\affiliation{School of Physics and Astronomy, University of Nottingham, NG 2RD, United Kingdom}
\affiliation{Ioffe Physical-Technical Institute, Russian Academy of Sciences, 194021 St.Petersburg, Russia}

\author{B. A. Glavin}
\affiliation{Lashkaryov Institute of Semiconductor Physics, 03028 Kyiv, Ukraine}

\author{V. I. Belotelov}
\affiliation{Lomonosov Moscow State University, 119991 Moscow, Russia}

\author{I. A. Akimov}
\affiliation{Experimentelle Physik 2, Technische Universit\"{a}t Dortmund, D-44227 Dortmund, Germany}
\affiliation{Ioffe Physical-Technical Institute, Russian Academy of Sciences, 194021 St.Petersburg, Russia}

\author{J. J\"{a}ger}
\affiliation{Experimentelle Physik 2, Technische Universit\"{a}t Dortmund, D-44227 Dortmund, Germany}

\author{S. Kasture}
\affiliation{Tata Institute of Fundamental Research, 400005 Mumbai, India}

\author{A. V. Gopal}
\affiliation{Tata Institute of Fundamental Research, 400005 Mumbai, India}

\author{A. S. Vengurlekar}
\affiliation{Tata Institute of Fundamental Research, 400005 Mumbai, India}

\author{D. R. Yakovlev}
\affiliation{Experimentelle Physik 2, Technische Universit\"{a}t Dortmund, D-44227 Dortmund, Germany}
\affiliation{Ioffe Physical-Technical Institute, Russian Academy of Sciences, 194021 St.Petersburg, Russia}

\author{A. J. Kent}
\affiliation{School of Physics and Astronomy, University of Nottingham, NG 2RD, United Kingdom}

\author{M. Bayer}
\affiliation{Experimentelle Physik 2, Technische Universit\"{a}t Dortmund, D-44227 Dortmund, Germany}
\affiliation{Ioffe Physical-Technical Institute, Russian Academy of Sciences, 194021 St.Petersburg, Russia}

\date{\today}

\begin{abstract}
Coherent sub-THz phonons incident on a gold grating that is deposited on a dielectric substrate undergo diffraction and thereby induce an alteration of the surface plasmon-polariton resonance. This results in efficient high-frequency modulation (up to 110 GHz) of the structure's reflectivity for visible light in the vicinity of the plasmon-polariton resonance. High modulation efficiency is achieved by designing a periodic nanostructure which provides both plasmon-polariton and phonon resonances. Our theoretical analysis shows that the dynamical alteration of the plasmon-polariton resonance is governed by modulation of the slit widths within the grating at the frequencies of higher-order phonon resonances.
\end{abstract}


\maketitle


Creating new devices based on plasmonic nanostructures (PNs) requires development of new physical concepts where the properties of plasmons and their interaction with photons may be controlled externally. Several methods of this "active plasmonics"\cite{Mac2009,Kni2011} were reported where the energy and propagation of plasmons were controlled by temperature\cite{Ler2005}, optical excitation\cite{Cas2010,Bai2009}, electric\cite{Has2012} and magnetic fields\cite{Tem2010,Bel2011,Bel2009}. In order to explore the properties of plasmons in nanodevices it is necessary to realize nondestructive control of plasmons on timescales far below 1 ns. In particular, such techniques could be employed in recently developed plasmon lasers (spasers), to enhance their functionality\cite{Ber2003,Nog2009}. Only then the advantage of plasmonics as compared traditional integrated electronics may be indeed exploited.\\ 
By now there are a number of works where ultrafast control of plasmons in PNs has been demonstrated using femtosecond optical excitation\cite{Poh2012,Rot2009} which possess  a number of undesirable side effects, like thermal heating or excitation of high-energy electron states. Besides modulation of the PN dielectric function, plasmonic states may be controlled by modulation of the geometrical parameters of the PN, like size of elements or distances between elements. This can be realized by applying uniaxial stress\cite{Chi2010,Chr2003} and, for dynamical modulation, acoustic waves may be used.  The feasibility of such an acoustic approach for the modulation of plasmonic properties has been already shown in a number of recent works, where THz phonons interact with the plasmon resonance in a very small noble metal particle\cite{Del1999,Hod1999}, or in periodic structures but in the frequency range up to \mbox{10 GHz}\cite{Im2004,Rup2010,Che2010}.\\
\begin{figure}[tbh!]
    \includegraphics[width=\linewidth]{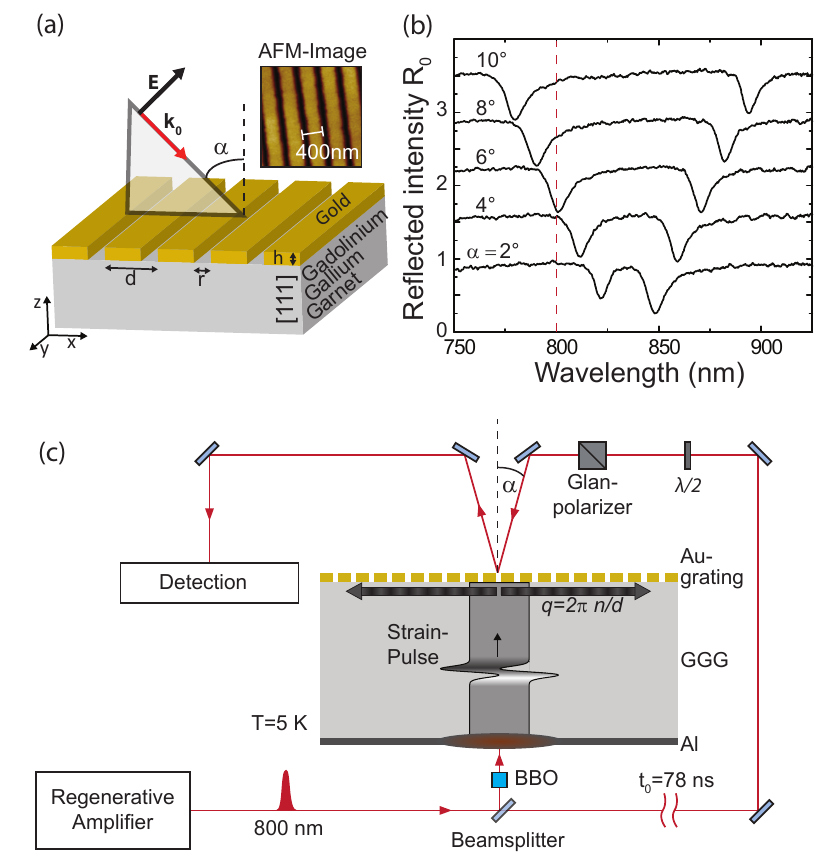}
    \caption{(a) The scheme of the sample and AFM image of the grating. (b) The reflectivity spectrum for p-polarized white light from the Au grating on GGG crystal measured for several incident angles; (c) The setup of the experiments with picosecond strain pulses. The horizontal arrows at the surface with the grating show schematically diffracted acoustic waves which modulate the surface plasmon-polaritons at the Au grating/GGG interface.}
    \label{Figur1}
\end{figure}
The aim of the present work is to realize an efficient modulation by sub-THz coherent phonons in a PN which has a narrow band plasmon-polariton resonance in the visible range.  The main difficulty in realizing such a task lies in the fact that the typical sizes determining the PN properties at optical frequencies are much larger than the wavelength $\Lambda$ of THz phonons (e.g. metal grating periods of \mbox{$d$=100 to 1000 nm} compared to \mbox{$\sim$10 nm} phonon wavelength). This difference results in a negligibly small modulation efficiency. In the present work we explore the diffraction of sub-THz phonons by a plasmonic grating. For such a concept the periodic character of the PN has two roles, given by two types of resonances: one is an optical, i.e. plasmon polariton resonance; and the second one is a phonon resonance related to high-order diffraction of near-surface longitudinal acoustic phonons.\\

The PN used in the present experiments is a gold (Au) grating fabricated on the (111) plane of a Gadolinium Gallium Garnet (GGG) substrate, shown in Fig.\ref{Figur1} (a). The grating period \mbox{$d$=400 nm}, thickness of Au stripes \mbox{$h$=80 nm} and width of slits $r$=50 nm were optimized to have distinct spectral lines in the specular optical reflectivity spectra shown in \mbox{Fig.\ref{Figur1} (b)}, recorded for several incidence angles $\alpha$ with the electrical field vector lying in the incidence plane (p-polarization). The dispersion of the spectral position of these lines versus $\alpha$ corresponds well to the solution of the momentum conservation law for the plasmonic wave at the GGG/Au interface\cite{Mai2007}:
\begin{eqnarray}
k_0\sqrt{\varepsilon_3}\sin{\alpha}=\beta+m\frac{2\pi}{d},
\label{eq:1}
\end{eqnarray}
where $k_0$ is the incident light wave number, $\varepsilon_1$, $\varepsilon_2$, $\varepsilon_3$ are the dielectric constants of gold, GGG, and air, respectively, $\beta=k_0\sqrt{\varepsilon_1\cdot\varepsilon_2/(\varepsilon_1+\varepsilon_2)}$, and $m$ is an integer. Comparison with the experimental reflection spectra indicates that the observed resonance features are due to excitation of the first-order surface-plasmon polaritons at the GGG/Au interface, i.e. $m$=1. Details of the optical properties of the studied PN may be found elsewhere\cite{Bel2011,Poh2012}.\\
The experimental scheme is shown in \mbox{Fig.\ref{Figur1} (c)}. A \mbox{50 nm} Al film was deposited on the GGG substrate surface opposite to the grating. It was optically excited by femtosecond pump pulses from regenerative amplifiers at 800 nm or 400 nm (frequency doubled) wavelengths. Two laser systems have been used: one with pulse duration 150 fs at a repetition rate 100 kHz and another with 60 fs pulses at 5 kHz repetition rate. The pump laser beam was focused to a spot with $\sim$100 $\mu$m diameter and the maximum excitation density was \mbox{W$\sim$10 mJ/cm$^2$}. The optical excitation of the Al film results in injection of a bipolar strain pulse into the GGG substrate, which has an amplitude up to 10$^{-3}$ and $\sim$10 ps duration\cite{Tho1986,Wri1994}.\\ 
The strain pulse, which corresponds to an acoustic wavepacket with sub-THz frequencies, propagates through the 500 $\mu$m thick GGG substrate at the velocity of longitudinal sound \mbox{6400 m/s} and reaches the grating after \mbox{$t_0=78$ ns}. There the strain pulse excites a number of elastic modes with wavevectors \textbf{q} parallel to the surface of the grating. There are two types of excited modes which are relevant for our analysis. The first type of modes originates from pure surface acoustic waves of GGG covered by the gold film. The zone-center waves are coupled to the normal incidence bulk waves. For our structure, we estimate the frequency of such waves to be in the range of 5 to 7.5 GHz. Similar modes were observed in previous acousto-optical experiments\cite{Ant2002,Che2010}. The second type of modes, shown schematically by the horizontal arrows in \mbox{Fig.\ref{Figur1} (c)}, has a bulk nature and propagates in GGG after diffraction from the periodic grating structure. Among these modes there are longitudinal acoustic (LA) waves propagating in the bulk of GGG with \textbf{q} parallel to the surface of the grating and perpendicular to the Au stripes, i.e. \textbf{q}$\|$\textbf{x}. These near-surface modes possess the properties of bulk LA waves in GGG and govern the most interesting observations in the present work.\\
Modes of both types (i.e. surface and near-surface modes) propagate along the Au grating/GGG interface and induce coherent displacements of the grating elements, thereby modulating its geometric parameters. Such modulation can be detected by monitoring the intensity $I$ of a reflected optical probe pulse with 800 nm wavelength, originating from the same laser and delayed relative to the pump pulse by \mbox{$t_0\approx$78 ns}, corresponding to the time after which the picosecond strain pulse hits the grating structure. In our experiments the probe beam was focused on the grating with a spot diameter of \mbox{30 $\mu$m} and an energy density per pulse of less than \mbox{0.05 mJ/cm$^2$}.\\

\begin{figure}[b!]
    \includegraphics[width=\linewidth]{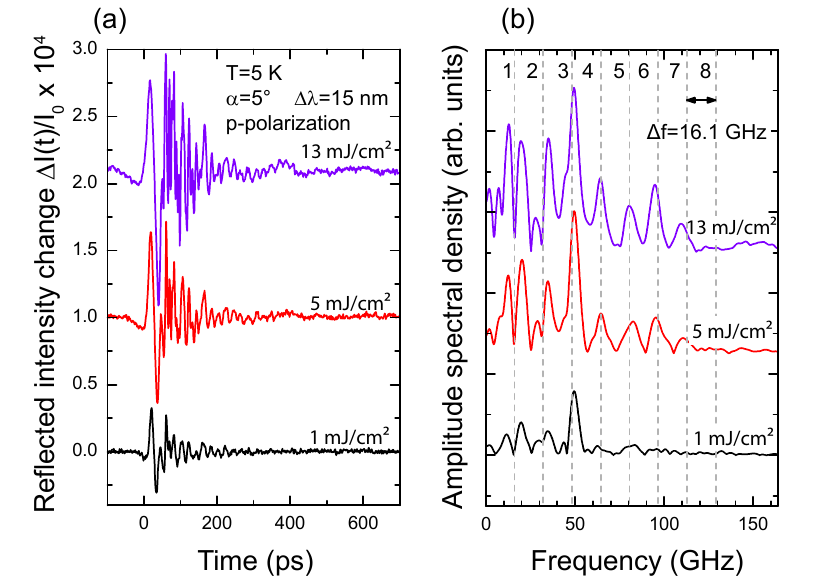}
    \caption{(a) The temporal evolutions of the reflected intensity changes induced by the picosecond strain pulse measured for three pump excitation densities $W$. The probing spectral width was \mbox{$\Delta\lambda$=15 nm} and the incident angle $\alpha=5^o$. (b) The amplitude fast Fourier transforms of the corresponding temporal traces is shown in (a). The vertical dashed lines indicate the frequencies expected for the acoustic modes diffracted by the Au grating and propagating parallel to surface in the GGG substrate [see Eq. (2)]. The integers at the top of the panel indicate the diffraction order $n$.}
    \label{Figur2}
\end{figure}

The experimental results obtained at cryogenic temperatures ($T$=5 K) are shown in Fig.\ref{Figur2}. Figure \ref{Figur2} (a) shows the temporal evolutions of the intensity changes $\Delta I(t)/I_0=(I(t)-I_0)/I_0$ for three pump excitation densities $W$ ($t$ is the time between the arrival of the strain pulse at the GGG/Au grating interface and the probe pulse, and $I_0$ is the stationary intensity of the reflected probe beam without the strain pulse). Complicated oscillatory behaviors of $\Delta I(t)/I_0$ are observed in all three curves. The exact evolution of $\Delta I(t)/I_0$ depends on $W$. In particular, the changes in reflectivity start earlier for high $W$ which is in agreement with appearance of nonlinear phenomena during strain pulse propagation for \mbox{$W>$1 mJ/cm$^2\ $}\cite{Hao2001}. The absolute changes of reflected intensity reach $2\times10^{-4}$ for the highest $W$ and depend also on the incidence angle and polarization of the probe beam. The curves in \mbox{Fig.\ref{Figur2} (a)} are obtained for $p$-polarized light at $\alpha=5^o$ when the central wavelength of the probe beam lies at 800 nm, i.e. on the short-wavelength flank of the plasmon-polariton resonance [see \mbox{Fig.\ref{Figur1} (b)}]. The signal is zero for probing by $s$-polarized light (electric field perpendicular to the incidence plane) for which no plasmon-polariton peaks are observed at any $\alpha$, due to the boundary conditions. Also, $\Delta I(t)/I_0=0$ for both polarizations for $\alpha=2^o$, because in this case the spectral overlap between the polariton and the probe spectral lines is negligible. Thus we may conclude that the observed signal $\Delta I(t)/I_0$ is exclusively resulting from modulation of the surface plasmon-polariton resonance by the elastic waves excited by the picosecond strain pulse.\\ 
	The amplitude spectra of the modulation signals, obtained by fast Fourier transformation of the measured $\Delta I(t)/I_0$, are shown in \mbox{Fig.\ref{Figur2} (b)} for the same $W$  as in \mbox{Fig.\ref{Figur2} (a)}. For the highest $W$ (top curve in \mbox{Fig.\ref{Figur2} (b)}) the spectrum spreads up to \mbox{$f$=110 GHz}. The spectrum consists of a number of peaks whose frequencies do not depend on $W$. For lower $W$, on the other hand, the high frequency peaks are less pronounced compared to high $W$ values.\\
	The remarkable result of \mbox{Fig.\ref{Figur2} (b)} is the observation of an almost equidistant spectral separation between the spectral lines for \mbox{$f>30$ GHz}. The central frequencies of these spectral lines are well described by
\begin{eqnarray} 
	f_n=\frac{n s}{d},
	\label{eq:2}
\end{eqnarray}
where $n$ is an integer and \mbox{$s$=6440 m/s} is close to the LA velocity for waves propagating in GGG along the (111) plane\cite{Kit1985}. The appropriateness of describing the modulation frequencies by Eq.(2) leads us to the explanation that the   components in the modulation spectrum are governed by the LA near-surface waves, which are diffracted by the grating and propagate in GGG with the sound velocity $s$ along the $\textbf{x}$ direction. The integer $n$ corresponds to the interference order of the diffracted waves. In other words, the Au grating plays the role of a diffraction grating for acoustic waves in addition to the plasmon-photon coupling that it provides. The diffracted acoustic waves propagating along the GGG/grating interface initiates the sub-THz modulation of the reflected light intensity in the spectral region of the surface-plasmon-polariton resonance. The spectral lines which would correspond to transverse acoustic (TA) phonons are not observed experimentally. If they were present, they would be easily recognized due to the smaller TA sound velocity. We attribute the absence of TA peaks to the mechanical boundary conditions which control the transformation of the incident LA strain pulse into diffracted TA waves. It is worth to mention that bulk to surface transformation of low frequency acoustic waves was shown earlier\cite{Hum1969,Yam1976}.\\
\begin{figure}[b!]
    \includegraphics[width=\linewidth]{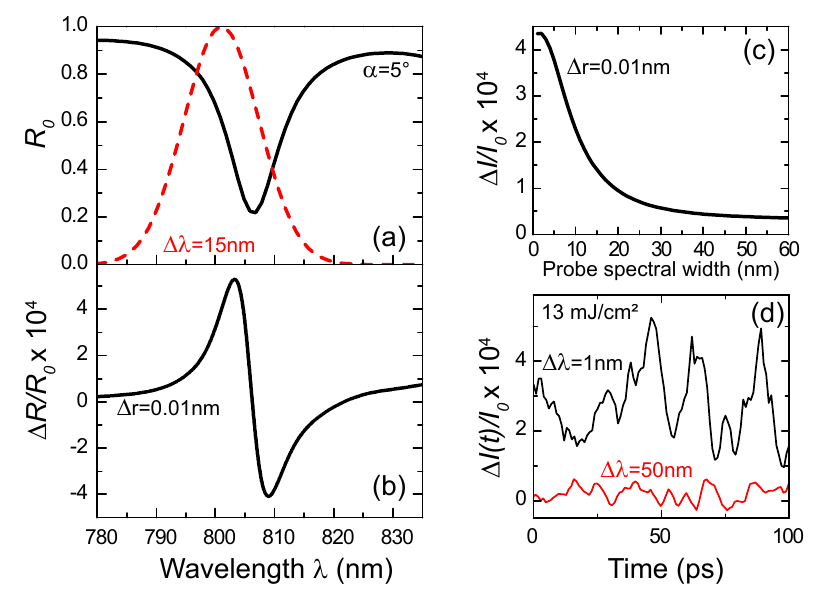}
    \caption{(a) The calculated reflectivity spectrum (solid line) and spectrum of the probe pulse with the spectral width \mbox{$\Delta\lambda$=15 nm} (dashed line); (b) The calculated differential spectrum $\Delta R/R_0$  when the width of the slit in the Au grating changes by \mbox{$\Delta r$=0.01 nm}; (c) The calculated dependence of the relative reflected intensity changes   for a probe pulse centered at 800 nm on the spectral width $\Delta\lambda$; (d) The measured 100 ps fragments of the reflectivity signals for probing with \mbox{$\Delta\lambda$=1 nm} (upper curve) and \mbox{$\Delta\lambda$=50 nm} (lower curve). In all curves the central wavelength of the probing beam was \mbox{$\lambda$=800 nm}, the incident angle was $\alpha=5^o$, and the temperature was $T$=5 K.}
    \label{Figur3}
\end{figure}
The widths of the spectral lines are almost independent on $n$ up to $n$=7. Such behavior is similar to optical diffraction on gratings where the spectral width for a certain angle does not depend on the diffraction order but is governed by the number $N$ of coherently interfering grating periods\cite{Bor1999}. From the experimental data, we estimate $N\approx 3$ for the number of "active" grating periods, corresponding to a lateral size of more than 1 $\mu$m. This number characterizes the lateral length across which coherence between the waves is conserved. Although we cannot address this effect quantitatively, we note that likely it is determined by the deviation of the acoustic field from periodicity in the lateral direction due to disorder of the grating.\\
Now we present the theoretical analysis of the coupling between the surface plasmon-polaritons and the near surface acoustic waves, which governs the measured optical response at high frequencies (\mbox{$f>30$ GHz}). The elasticity equations show that the lateral displacement-profile is an odd function of $x$ with respect to the middle of the grating stripes\cite{Gla1981}, provided that the grating profile is an even function of $x$. Therefore, we may assume that the main contribution to $\Delta I(t)/I_0$ comes from the coherent modulation of the slit width $r$, while the period $d$ of the grating remains unperturbed. The influence of $r$ on the reflection spectrum can be calculated by the rigorous coupled waves analysis (RCWA) which allows us to model the optical properties of periodic multilayered structures\cite{Moh1995,Li2003}.\\
The calculated, unperturbed reflectivity spectrum $R_0$ for $\alpha=5^o$ along with the probe spectrum is shown in \mbox{Fig.\ref{Figur3} (a)}. At $\alpha=5^o$ the probe central wavelength is on the shorter-wavelength wing of the $R_0$ spectrum. The RCWA modeling shows that small changes $\Delta r$ of the slit width cause a shift of the plasmonic resonance in the reflectivity spectrum. That is why the change of the reflectivity spectrum $\Delta R(\lambda)/R_0(\lambda)=(R(\lambda)-R_0(\lambda))/R_0(\lambda)$ has positive and negative peaks corresponding to the wings of the plasmonic spectral line in reflection. A calculated spectrum $\Delta R/R_0$ for \mbox{$\Delta r$=0.01 nm} is shown in \mbox{Fig.\ref{Figur3} (b)}. The measured signal $\Delta I/I_0$ integrated over the probe spectral width is sensitive to the spectral bandwidth of the probe pulse $\Delta\lambda$ (\mbox{Fig.\ref{Figur3} (c)}). The largest signal is expected for the case when the probe spectrum overlaps with the positive or negative parts of the $\Delta R/R_0$ spectrum only. This implies that there is some optimal incidence angle at which the central wavelength of the probe is located on the spectral wing of the plasmonic resonance and $\Delta\lambda$ is such that only one peak of $\Delta R/R_0$ acts.\\ 
The experimentally measured amplitude  $\Delta I/I_0\sim10^{-4}$ is in good agreement with theoretical calculations, if we assume the lateral displacement to have a magnitude of \mbox{$\Delta r$=0.01 nm}, resulting in a corresponding change of the slit width. For $n$=7, this number corresponds to strain of about 10$^{-3}$. Taking into account, that the amplitude of strain in the incident pulse and its spectral width are $\sim10^{-3}$ and 100 GHz, respectively, we may conclude that the grating gives rise to considerable local enhancement of the acoustic field in the relatively narrow spectral bands near $f_n$.\\
In order to confirm the calculated dependence of the modulation amplitude on $\Delta\lambda$ shown in \mbox{Fig.\ref{Figur3} (c)} we have performed experiments with three different values of $\Delta\lambda$. The experimental results shown in Fig.\ref{Figur2} are obtained for \mbox{$\Delta\lambda$=15 nm}. The signals   measured for \mbox{$\Delta\lambda$=1 nm} and \mbox{$\Delta\lambda$=50 nm} are presented in \mbox{Fig.\ref{Figur3} (d)}. The amplitudes of high-frequency oscillations are 5 times higher for \mbox{$\Delta\lambda$=1 nm} than for \mbox{$\Delta\lambda$=50 nm}. This is in reasonable agreement with theory predicting an order of magnitude difference. This agreement supports our assumption that the coherent modulation of the slit widths is the main mechanism for coupling the surface plasmon-polaritons with the LA near-surface acoustic waves.\\

\begin{figure}[t!]
    \includegraphics[width=\linewidth]{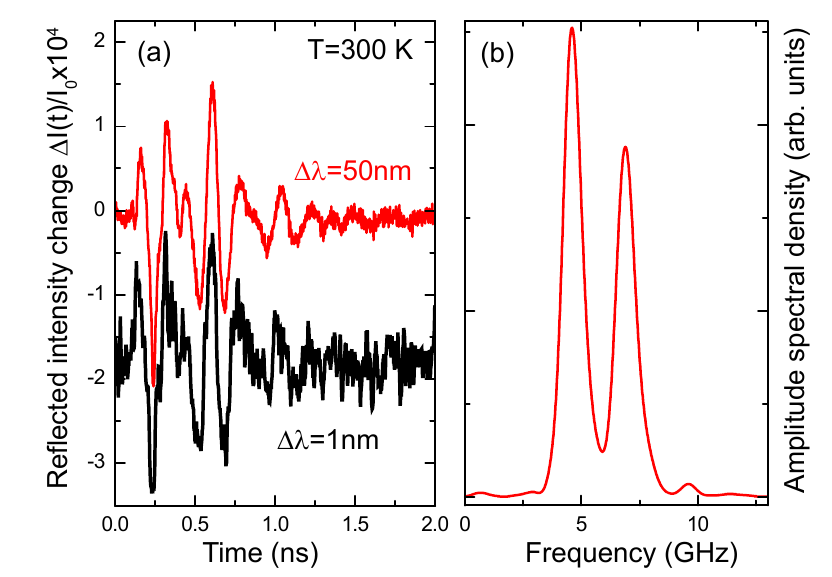}
    \caption{(a) The time evolution of the reflected intensity at room temperature for \mbox{$\Delta\lambda$=50 nm} (upper curve) and \mbox{$\Delta\lambda$=1 nm} (lower curve), (\mbox{$\lambda$=800 nm}, $\alpha=5^o$, and \mbox{$W\sim10$ mJ/cm$^2$}). (b) Corresponding fast Fourier transform for \mbox{$\Delta\lambda$=50 nm}.}
    \label{Figur4}
\end{figure}

Now we turn to discussing the low frequency (\mbox{$f<30$ GHz}) part of the modulation spectrum, which consists of a series of spectral lines with central frequencies not described by Eq. (2), see \mbox{Fig.\ref{Figur2} (b)}. For instance, the spectrum has a dip at \mbox{$f=16.1$ GHz} instead of the expected peak related to the fundamental ($n$=1) LA near-surface mode. We attribute the spectral lines in the spectral region \mbox{$f<30$ GHz} to various surface modes excited in the Au grating as observed also in earlier works\cite{Ant2002,Che2010}. The exact spectrum of the modulated signals depends on the dispersion of these modes and their interaction with bulk modes in GGG. For instance the dip at \mbox{$f=16.1$ GHz} may be due to a Fano resonance originating from interaction of the surface modes in the Au grating and LA near-surface modes in GGG.\\ 
We performed also experiments at a temperature of 300 K when low frequency components are fully dominant, because high-frequency oscillations in $\Delta I(t)/I_0$ are damped completely due to the strong attenuation of sub-THz LA waves during their propagation through the GGG substrate\cite{Krz1984}. The results are presented in \mbox{Fig.\ref{Figur4}}. The signals   do not depend on $\Delta\lambda$ which means that the interaction mechanism for the low frequency modes with the polariton resonance is different from the one described above for \mbox{$f>30$ GHz}, where the modulation of the grating slit width is considered as the main perturbation. Comprehensive analysis of the surface modes is beyond the scope of the present paper where we concentrate mostly on the high-frequency part of the modulation spectrum.\\
In conclusion, we have demonstrated that coherent phonons possess diffraction on the plasmonic grating and this results in efficient modulation of a plasmon-polariton resonance at frequencies \mbox{$\sim$100 GHz}.


%




\begin{acknowledgments}
We acknowledge financial support by the Deutsche Forschungsgemeinschaft (project BA 1549/14-1), the Russian Foundation for Basic Research (RFBR), the Russian Presidential Grant MK-3123.2011.2, and the Federal Targeted Program "Scientific and Scientific-Pedagogical Personnel of the Innovative Russia" (Governmental Contract No. 16.740.11.0577). Financial support for the work in Nottinghan was provided by the UK Engineering and Physical Sciences Research Council. The authors like to thank Prof. V. Gusev for the discussion of the properties of surface acoustic waves.
\end{acknowledgments}

\bibliography{plasmonics_bibtex}

\end{document}